# ISLANDING MICROGRID WITH SOLAR AND FUEL CELL FOR PROVIDING LOAD

*Mohsen Pasandi[1]*

*Master's student in the Electrical Engineering department, Clausthal University of Technology, Germany [1]*

*Mohsen.pasandi@tu-clausthal.de[1]*

**Abstract:**
*Solar and Fuel cell energy resources are two major of Renewable Energy Resources (RES) which is using to power electrical grids. They can provide clean energy and help to reduce greenhouse emissions. RES is changing the structure of electrical systems day by day. They have provided to supply the loads and have distributed systems. The concept of Microgrid has been defined by connecting distributed generation sources and load which can run in islanding and interconnected modes. In this research, solar and fuel cell are used as a source for providing variable loads. They try to deliver defined power to the load. The solar system created by one diode corresponding circuit is given and a Perturbation and observation (P&O) approach is applied to obtain full power from the solar panels. Direct Methanol Fuel Cell Model (DFMC) is another renewable energy resource that is utilized in this microgrid to provide the load. This model has included two Gibbs reactors that considered for the anode and cathode respectively and a splitter among the anode and cathode. The droop controller is used to control the injected power from the resources to the loads. The effectiveness of the designed microgrid is justified in MATLAB/SIMULINK environment.*

## I. OVERVIEW INTRODUCTION

A microgrid is an integration of Distributed Generation Resources (DGR) such as solar energy, wind energy small hydro, biomass, biogas, geothermal, and fuel cell. In other definition, it is a small-scale power grid that can operate independently or collaboratively with the utility system. To run microgrids to work properly, a switch should be able to disconnect microgrid with the upstream grid. In this case, the microgrid will work in islanding mode. In the islanding situation, loads will be fed by DGR. The voltage and frequency should be maintained in the acceptable range in this condition [1] – [2].

Because, all the required energy for microgrid comes from DGR, control, and programming of these resources to have proper performance is an assist. Power electronic devices give an option for controlling DG's. By designing the appropriate control strategy for connected inverters, they can provide active and reactive required powers during islanded operation and interconnected mode [2]. By considering all afore-mentioned issues, designing control strategies for inverters to work properly in both operation modes has the highest priority for researchers in the field. Two prominent sources that have a high potential for proving loads in microgrids are PV farms and fuel cells. The use of renewable energy sources in recent years has been growing rapidly, but among them, the solar energy source is one of the available and promising sources. Designers before installing the PV systems require a precise model to achieve the best result to simulate. The model that considers all standard and operational conditions. Modeling the PV system is one of the most important parts of the power electronics challenges, such that the electrical behavior of the PV system can be Combined with these elements. In this project, a comprehensive, authoritative and user-friendly model of photovoltaic (PV) arrays in mathematical model And Maximum Power Point Tracker (MPPT) is presented. This paradigm collects the amount of radiation and temperature as the input parameters and provides Power-Voltage (P-V) and Current–Voltage (I-V) diagrams as the output result. By using the Maximum Power Point Tracker (MPPT) maximum output power can be achieved [3] – [4]. Several great advantages such as a high level of energy density, low cost and easy to work within distribution make DMFC an attractive power source [5] – [6]. Thus, for having enough energy and long operating time, they do not need to connect to an external power source. Also, they can be re-fueled which this characteristic makes them quite appropriate for using them in remote areas. This technology has several problems which should be solved. Obstacles like high cost and quick degradation [8] – [9].

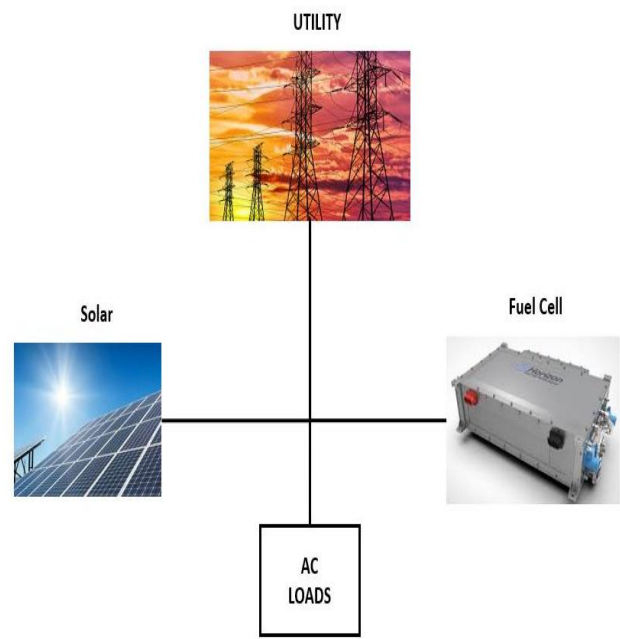

FIGURE (1):
THE DEFINED MICROGRID

Figure (1) shows the defined microgrid which is included in a PV system and a fuel cell system. A lot of researches are conducted on microgrids to improve their capabilities such as hybrid microgrids, power quality issues, optimization, prices and connecting electric vehicles to them and some other aspects of microgrid need to be considered such as forecasting or numerical methods. Different software such as LTSpice is also a good tool [10]- [15]. In this paper, a microgrid is included solar and fuel cells as sources and loads. These two sources supply the load by receiving orders to their inverters by the Droop controller. The structures of those sources and the designed controller are explained completely. The effectiveness of the designed microgrid is justified in MATLAB/SIMULINK environment.

## II. PV SYSTEM

PV cell simulation includes obtaining characteristic (P-V) and (I-V) curves. The goal of this work will be adapting the characteristics curves of the Simulation Model with the characteristic curve of the actual Cell under various environmental conditions. The recent method can be utilized as an equivalent electrical circuit which will be based on a diode model. Researchers have provided several models for considering solar cells which the simplest model which has been introduced is a single diode model. This model has an autonomous current source, which is coupled in parallel to a diode and is shown in Figure (1). The mathematical equations related to this model can be found in the below equation and the detailed has been expressed at [3] – [4].

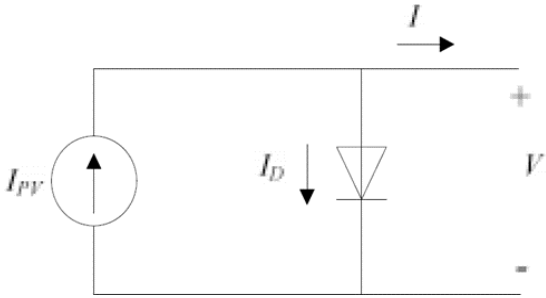

FIGURE (2):
THE IDEAL SOLAR CELL MODEL

$$I = I_{pv} - I_D \quad (1)$$

$$I = I_{PV} - I_o \left[ exp\left[\frac{V}{\alpha V_T}\right] \right] \quad (2)$$

$$V_T = \frac{KT}{q} * nl * N_{cell} \quad (3)$$

Solar cells have a characteristic that varies like a variable resistance. It means that on the following diagrams Figure (3) for each voltage, it adjusts its current. An inverse situation, if the current is set its voltage varies according to the diagram. Therefore, in modeling, the voltage or current is by MPPT, the maximum possible electrical power of an array (PV) is achieved. Accomplishing this goal will be by pushing the solar system to work at or close to the peak point of power in the different conditions for PV panels such as irradiance, temperature, and load. always received at the output feedback to determine the operating point of the solar cell. And the task of MPPT is to adjust the working point to the knee point because it has the highest power production of the cell [5] – [6].

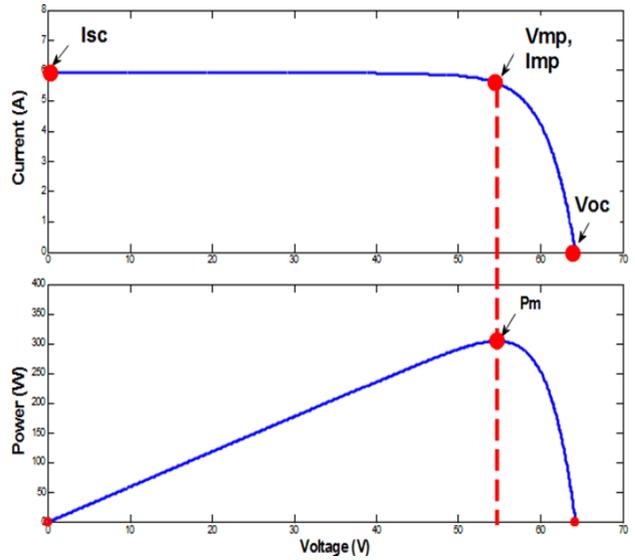

FIGURE (3):
THE (P-V) AND (I-V) CURVES [7]

This DC-DC converter changes the voltage level to achieve maximum output power. Changing voltage value is implemented by changing the duty cycle. There are three methods to do this:

1. Incremental conductance method
2. Perturbation and observation (P&O) method
3. Fractional open-circuit voltage method

In this paper, the (P&O) method has been used. This method tries to perturb the operating voltage to achieve maximum power by being in MPP. In recent years lots of research have been conducted to improve this method. In this paper, a fundamental P&O for reaching MPPT is demonstrated below. Mechanism of this algorithm works as follow:

1. Making a change in operating point
2. Calculating P and V
3. Evaluating new P with earlier one according to V.

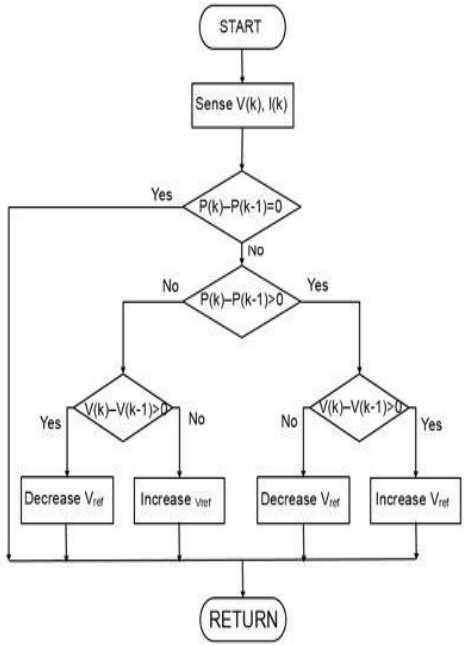

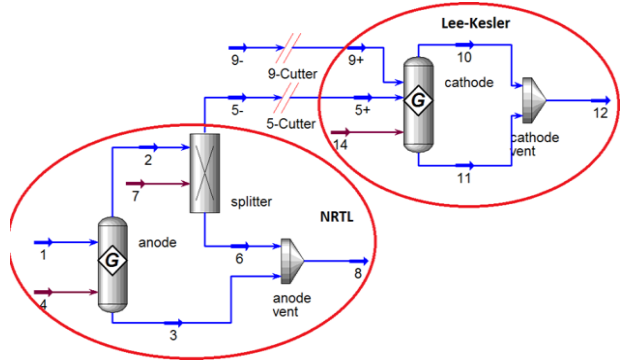

FIGURE (5):
Direct Methanol Fuel Cell Model [5]

FIGURE (4):
P&O ALGORITHM

Base on this is determined that slop is positive or negative and is it approaching MPP [3] – [7].

## III. DIRECT METHANOL FUEL CELLS

Figure (5) shows a diagram of the Direct Methanol Fuel Cell Model. This model has included two Gibbs reactors that considered for the anode and cathode respectively and a splitter among them [12]. After considering below chemical equations maximum cell voltage and maximum current density can be calculated based on the Nernst equation, Eq. 6 and Faraday's law, Eq. 7 respectively.

$$E_{max} = \frac{RT}{2F} \ln \frac{P_{H2\ an}}{P_{H2\ cat}} \quad (4)$$

$$I_{max} = \frac{2\ F_{nH2}}{A} \quad (5)$$

The model has been introduced by authors in [13]. Figure (6) shows the equivalent electrical circuit for DCMF. The maximum cell voltage characteristic is represented by Emax for an electrochemical reaction. Therefore, Cell voltage (E) will be obtained after considering for ohmic losses.

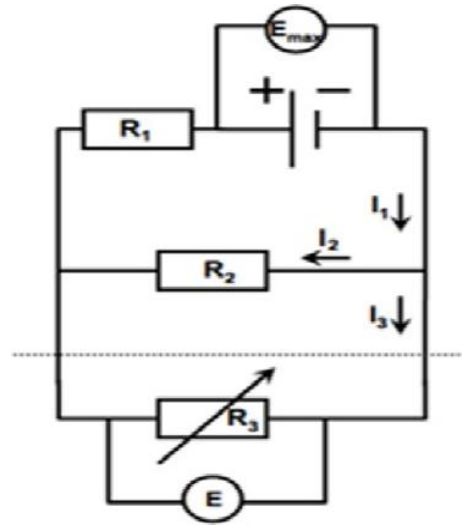

FIGURE (7): AN EQUIVALENT ELECTRICAL CIRCUIT OF A DIRECT METHANOL FUEL CELL [6]

I1 represents electric current which has been generated in a fuel cell. The methanol crossover and other losses will be presented by I2. Resistors R1 and R2 represented cell's ionic and electrical resistance respectively. The load (R3) was supplied by electric current I3. It was changed from zero to one resulting in current density change from zero to Imax. The number of hydrogen particles produced in the methanol oxidation reaction was directly proportional to the electrical current according to Faraday's law. Cell voltage (EDMFC) was calculated from Eq. (8). Eq. (8) was derived from the equivalent electrical circuit.

$$E_{DMFC} = \frac{E_{max} - r_1 \cdot \eta_{H2} \cdot i_{max}}{\frac{r_1}{r_2} \cdot (1 - \eta_{H2}) + 1} \quad (6)$$

The presented model was used to establish the polarization curves for a methanol concentration of 3 mole/dm3 [6].

## IV. DROOP CONTROL

The droop control concept has been achieved by the below equations where the inverter voltage is $E\angle\delta$. This voltage tries to deliver the produced power to the grid with the $V_o\angle 0^\circ$ voltage by $Z_o\angle\theta$ impedance. Both

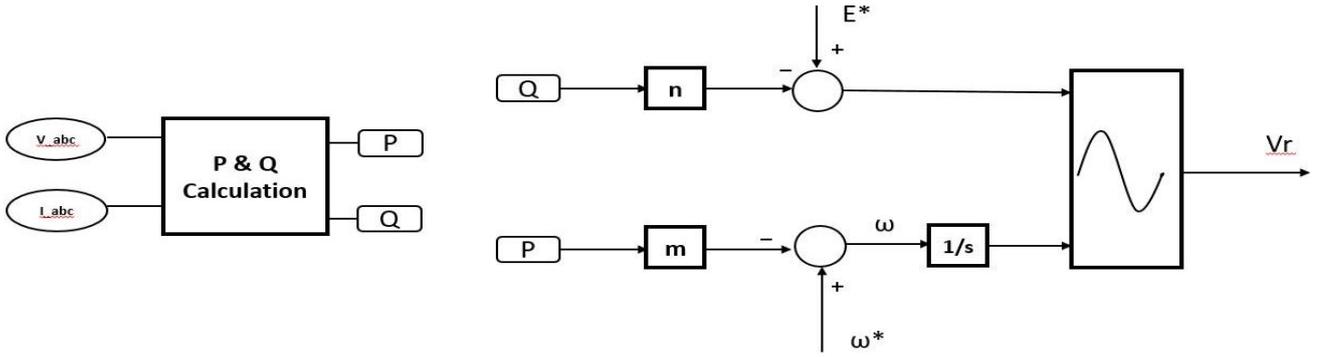

FIGURE (8):
DROOP CONTROLLER

active power P and reactive power Q obtained by the utility $V_o\angle 0^\circ$ created by [15] – [18] as

$$P = \left(\frac{EV_0}{Z_0}\cos\delta - \frac{V_0^2}{Z_0}\right)\cos\theta + \frac{EV_0}{Z_0}\sin\delta\sin\theta \quad (7)$$

$$Q = \left(\frac{EV_0}{Z_0}\cos\delta - \frac{V_0^2}{Z_0}\right)\sin\theta - \frac{EV_0}{Z_0}\sin\delta\cos\theta \quad (8)$$

Where the difference between the voltage phases among the two resources is shown by $\delta$. By considering the impedance of inductive $Z_o\angle\theta$ equations will change as

$$P = \frac{EV_0}{Z_0}\sin\delta \quad (9)$$

$$Q = \frac{EV_0}{Z_0}\cos\delta - \frac{V_0^2}{Z_0} \quad (10)$$

The Power angle is assumed as a small value. Thus,

$$E_i = E^* - n_i Q_i \quad (11)$$

$$\omega_i = \omega^* - n_i P_i \quad (12)$$

Where the setpoint voltage E* is named as is the frequency setpoint ω*, the rated voltage will be E, and will be named as the rated frequency ω the droop coefficients are considered as n and m which can be found base on the capacity of inverters and usually defined by the constraint of consumers [19] –[20]. Figure (8) demonstrates a view of the designed droop controller.

## I. SIMULATION RESULTS

The Identified microgrid in Figure (1) has been created in MATLAB/SIMULINK environment. The characteristics of microgrid, resources, and load are given in Table (1).

TABLE I
CHARACTERISTICS OF MICROGRID

| Characteristics | values |
|---|---|
| Voltage RMS | 320 (v) |
| Frequency | 5 Hz |
| Lf | 35 mh |
| CF | 3µf |
| FC Active Power | 8 KW |
| PV Active Power | 8 KW |
| FC Reactive Power | 1 Kvar |
| PV Reactive Power | 1 Kvar |
| LOAD | 16 KW, 2 Kvar |

Based on the given values in Table (1) fuel cell and PV systems should provide enough power to run the load. Figure (9) and Figure (10) show the simulation results for FC and PV productions. In the left side for Figure (9), Voltage, Current and Active and Reactive powers produced by designed and simulated PV and in the Right-side Voltage, Current and Active and Reactive powers produced by designed and simulated FC are given. Figure (10) shows the Voltage, Current and Active and Reactive powers for LOAD. It is proven that by the PV and FC sources and using a droop controller, the defined load is provided by good voltage and current.

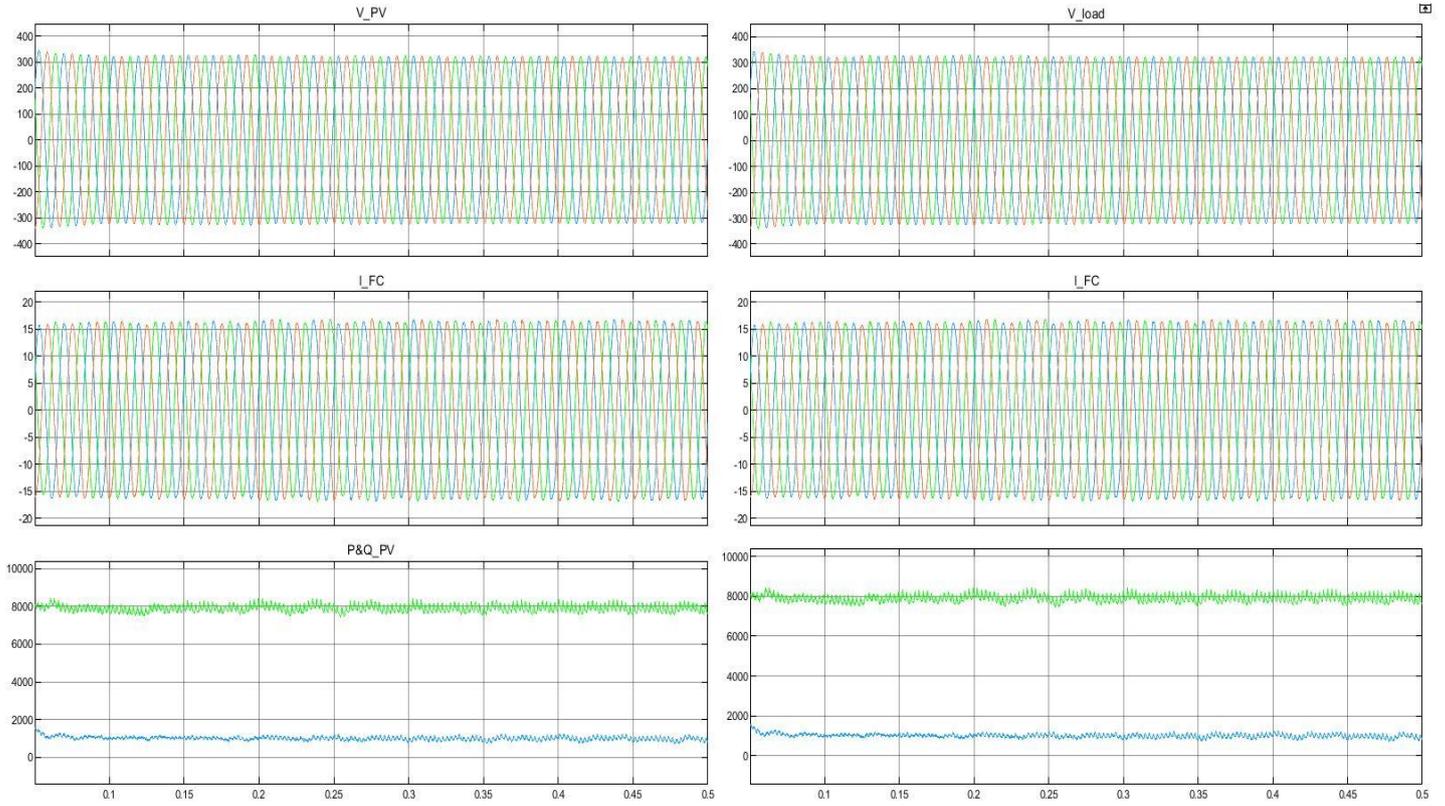

FIGURE (9):
Simulation Results: left figure (Voltage, Current and Active and Reactive powers for PV)
Right figure (Voltage, Current and Active and Reactive powers for PV)

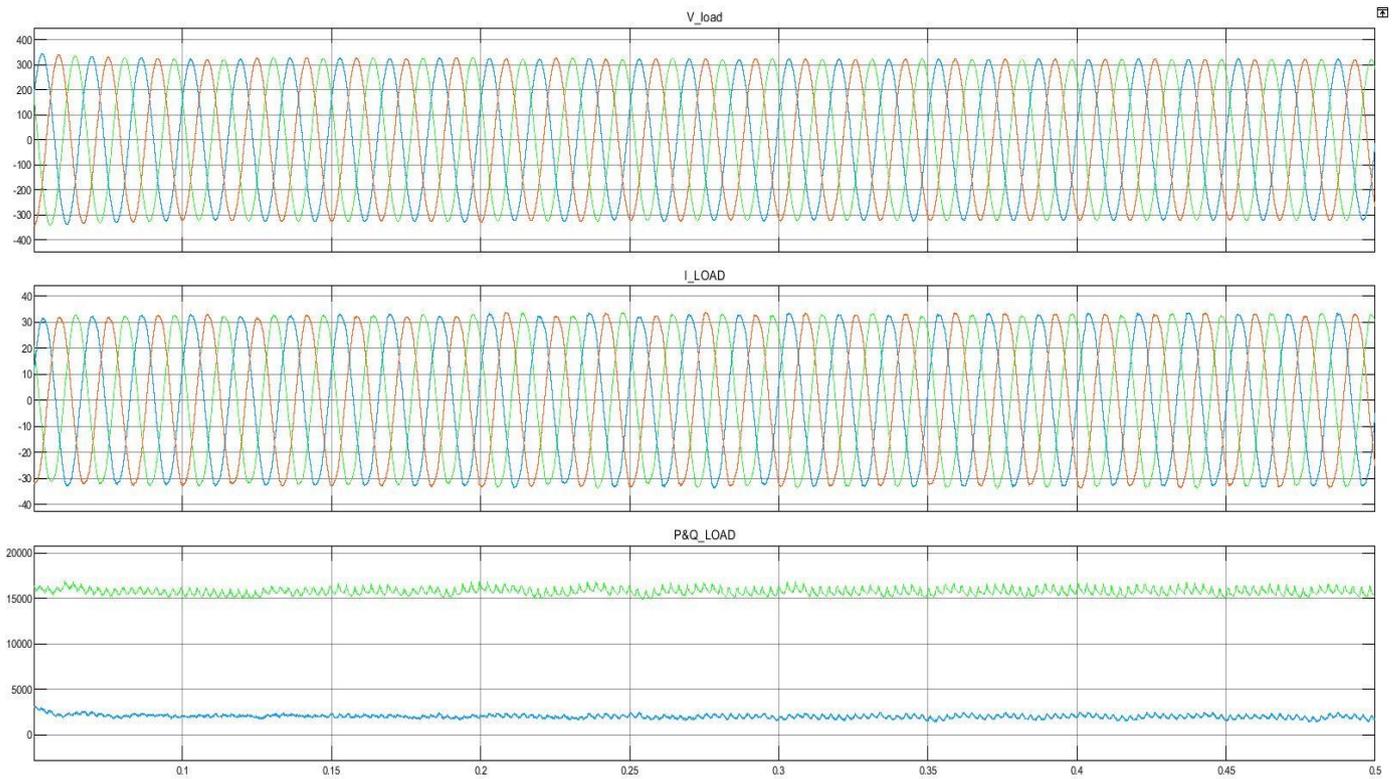

FIGURE (10):
Simulation Results for Voltage, Current and Active and Reactive powers for LOAD

## CONCLUSION

In this research, a microgrid with the integration of solar and fuel cell systems and the designed load. P&O algorithm was used to get maximum power from the PV system. The droop controller was explained shortly to clarify the system clearly. In the end, to verify the usefulness of designed microgrid, simulation results on MATLAB/SIMULINK environment were given.